\documentclass[aps,prl,reprint,superscriptaddress]{revtex4-1}
\usepackage{graphicx}
\begin{document}
\title{Phase Diagram and Soliton Picture of an Ideal Spin-Peierls Compound D-F$_{5}$PNN}
\author{Yuji Inagaki}
\email[inagaki.yuji.318@m.kyushu-u.ac.jp]{}
\author{Tatsuya Kawae}
\affiliation{Department of Applied Quantum Physics, Faculty of Engineering, Kyushu University, Fukuoka 819-0395, Japan}
\author{Naoko Sakai}
\author{Naoyuki Kawame}
\author{Takao Goto}
\author{Kunio Taguma}
\author{Jun Yamauchi}
\affiliation{Graduate School of Human and Environmental Studies, Kyoto University, Kyoto 606-8501, Japan}
\author{Yasuo Yoshida}
\affiliation{Institute for Solid State Physics, University of Tokyo, Kashiwa, Chiba 277-8581, Japan}
\author{Yutaka Fujii}
\affiliation{Research Center for Development of Far-Infrared Region, University of Fukui, Fukui 910-8507, Japan}
\author{Takashi Kambe}
\affiliation{Department of Physics, Okayama University, Okayama 700-8530, Japan}
\author{Yuko Hosokoshi}
\affiliation{Department of Physical Science, Osaka Prefecture University, Osaka 599-8531}
\author{B\'eatrice Grenier}
\affiliation{CEA-Grenoble, DRFMC/SPSMS/MDN, 17 rue des Martyrs, 38054 Grenoble cedex 9, France}
\author{Jean-Paul Boucher}
\affiliation{Laboratoire de Spectrom\'etrie Physique, Universit\'e J. Fourier, BP 87, 38402 Saint Martin d'H\'eres, France}

\begin{abstract}
We investigate magnetic properties of an $S$=1/2, quasi-one dimensional organic antiferromagnet, D-F$_{5}$PNN using magnetization measurements taken at temperatures as low as 0.5 K. Three distinct phases were observed consisting of uniform, dimerized (D), and incommensurate (I) spin structures in the magnetic field versus temperature plane, where a significant hysteresis appears between D-I transitions in the field scan measurements. A combination of magnon ($S$=1) and soliton ($S$=1/2) excitations have successfully reproduced the observed magnetic susceptibility. In addition, such excitations provide a reasonable interpretation of the temperature dependent electron spin resonance (ESR) spectra. By comparison with the theoretical study, we conclude that D-F$_{5}$PNN is an ideal compound for investigating　the spin-Peierls transition.
\end{abstract}
\maketitle

Research on low-dimensional magnetic systems is a major topic in solid state physics, because magneto-structural correlation brings about a variety of particular features in their ground state properties. The spin-Peierls (SP) transition was first proposed in the 1960 as a magnetic analogue of the Peierls transition in a one dimensional (1D) conductor.　　 
In the Peierls transition, an insulating ground state with a lattice distortion is stabilized due to an instability of the 1D chain. Similarly, in the SP transition, 1D chains with antiferromagnetically coupled spins dimerize spontaneously to lower the magnetic energy. This leads to the formation of a singlet ground state and the opening of an energy gap in the excitation spectrum below the transition temperature $T_{SP}$\cite{wbp1, *wbp2, *wbp3}. 

In the electronic counterpart, a modification of the lattice distortion is, in principle, achieved by changing the chemical potential and eventually filling of Fermion band. It should be noted that this process is difficult to control experimentally. However, the effect can be investigated by applying a realistic magnetic field to the SP system. Hence, considerable attention is focused on the shape of the applied magnetic field ($H$) versus temperature ($T$) phase diagram for the SP system\cite{h-t, cross}. Theoretically, three phases are expected to be stabilized and join at a Lifshitz point. At high temperatures for $T$ $>$ $T_{SP}$, a 1D Heisenberg chain forms a uniform structure, which is abbreviated as a U phase in this paper. A commensurate, dimerized (D) structure of a 1D chain appears when the temperature is decreased below $T_{SP}$. The application of a high magnetic field below $T_{SP}$ induces an incommensurate (I) phase, in which the wave vector describing the lattice distortion differs slightly from that of the D phase.

For the transition at the D-I boundary, Nakano and Fukuyama pointed out that the softening of soliton excitations plays an essential role \cite{nf}. The soliton, which is regarded as a domain wall or kink between two different patterns of bond alternations, has a net spin of 1/2. The creation energy of a soliton is considerably lower than that of a magnon so that a modulated, incommensurate structure is stabilized in a high magnetic field \cite{soliton, *soliton2}. Therefore, one of the crucial issues in the SP transition is whether the D-I boundary is understood by utilizing the soliton picture. 

First experimental evidence of the SP transition has been obtained for an organic compound, tetrathiafulvalinium bis-cis-(1,2-perfluoromethylethylene-1-2-dithiolate)-copper [TTF-BDT(Cu)] and its sister compound, TTF-BDT(Au). Intensive studies have revealed important aspects, such as the ordering nature at each boundary, the soliton picture and the high magnetic field incommensurate phase\cite{org1, org2, org3, org4, org5, org6, org7, org8}. However, because of the difficulty in preparing high quality organic crystals, the soliton excitations has not been evidenced in the thermodynamic quantities observed so far\cite{h-t, org5, org7, org8}. Following on from these pioneering works, in-depth research of the SP transition has been performed in an inorganic compound CuGeO$_{3}$\cite{cugeo}. It is, however, well known that CuGeO$_{3}$ is far from an ideal compound to study the SP transition because of the existence of sizable inter-chain coupling and intra-chain frustration.

In this paper, we report the results of magnetization and electron spin resonance (ESR) measurements down to $T$=0.5 K on an $S$=1/2, quasi 1D antiferromagnet D-F$_{5}$PNN, which is the deuterium version of an organic compound, pentafluorophenyl-nitronyl-nitroxide F$_{5}$PNN. The low-$T$ magnetic properties of F$_{5}$PNN are well understood as a $S$ = 1/2 1D Heisenberg AF alternating chain system\cite{f5pnn1, *f5pnn2, *f5pnn3, *f5pnn4, *f5pnn5}, whereas, in this work, that of D-F$_{5}$PNN, is found to be an ideal compound to investigate the SP transition. The magnetism of D-F$_{5}$PNN originates from organic nitronylnitroxide radicals, which indicates that the present compound is a good candidate for investigating the SP transition without magnetic anisotropy. In addition D-F$_{5}$PNN has advantages that high quality single crystal is available and the critical field of the transition is within an experimentally accessible range. Through detailed measurements, we describe here the precise $H$-$T$ phase diagram, which is in good agreement with that predicted by the theory. The spin gap estimated from the low-$T$ susceptibility is reasonably justified by considering the presence of low-energy soliton excitations. This is further confirmed by ESR spectroscopy.

Before showing the results, prior studies on D-F$_{5}$PNN are summarized here. A strong magneto-elastic effect is already reported by Can\'evet $et$ $al$.\cite{d-f5pnn-nd}. Using neutron scattering measurements, they observed a structural phase transition from the U phase to the D phase below about 1 K. In addition, the U phase was revived from the D phase by applying a magnetic field. A field-induced magnetic ordering (FIMO) phase was detected by specific heat measurements above 2 T below $T$ $\sim$ 0.26 K\cite{d-f5pnn-ct}. These experimental facts will be referred to later. 

Details of the sample preparation and the structural analysis of a D-F$_{5}$PNN, single crystal used in the present study will be published elsewhere\cite{sakai}. Magnetization ($M$) measurements were performed down to $T$=0.49 K and up to $H$=7 T using a commercial SQUID magnetometer, magnetic property measurement system (MPMS), equipped with a homemade $^{3}$He insert\cite{3he-m}. ESR spectra were obtained by the use of a Bruker X-band ESR system equipped with a $^{3}$He cryostat\cite{3he-esr}.

\begin{figure}
\includegraphics[width=60mm]{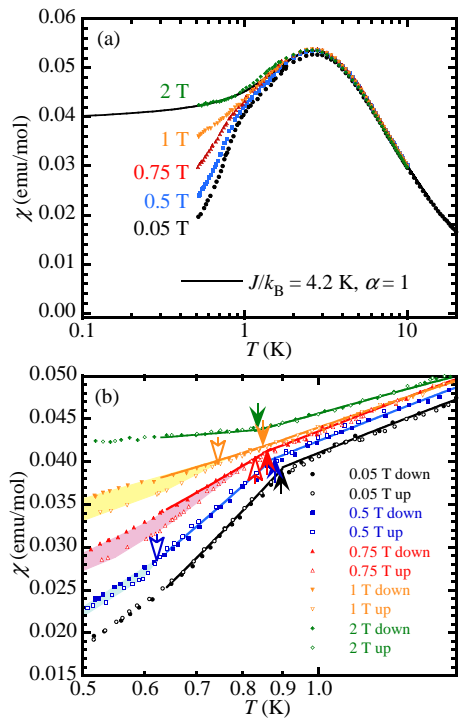}
\caption{\label{f1} (a) $\chi$($T$) taken in a $T$ decreasing process under various magnetic fields. Theoretical curve for the $S$=1/2 AF uniform chain ($\alpha$=1) with $J$/$k_{B}$ = 4.2 K is represented by a solid line.
(b) Low-$T$ part of $\chi$($T$) for increasing $T$ (open symbols) and decreasing $T$ (solid symbols) processes were recorded after zero-field cooling and under field, respectively. 
}
\end{figure}

Figure~\ref{f1}(a) shows the $T$ dependence of magnetic susceptibility $\chi$($T$) under various $H$ parallel to the chain direction. 
All data are taken in a decreasing $T$ process. At the lowest field of $H$ = 0.05 T, $\chi$($T$) shows a steep decline toward zero magnetic susceptibility below $T$ = 0.9 K, which is assigned as $T_{SP}$. This decline is caused by the SP transition, at which the gapped ground state is progressed. 
Above $T$ = 1 K, $\chi$($T$) with a broad maximum at around 2.7 K is well-reproduced by a numerical calculation for an $S$ = 1/2 AF Heisenberg linear chain model with the interaction $J$/k$_{B}$=4.2 K, as shown by the solid line in Fig.~\ref{f1}(a)\cite{johnston}. Here, we used the value of $g$ = 2.006 for the chain direction measured by X-band ESR at room $T$. However, it can be seen here that $\chi$($T$) deviates from the calculation even above $T_{SP}$. This may be attributed to the emergence of a pseudo gap due to the structural fluctuation, which is expected by the theory and is confirmed in several other compounds \cite{pseudo-gap1, pseudo-gap2, pseudo-gap3}. 

When the magnetic field is applied, the decline of $\chi$($T$) below $T_{SP}$ is reduced, which is understood by the suppression of the energy gap. In practice, $\chi$($T$) approximately follows the calculation in the entire $T$ range at $H$=2 T, where the gap is observed to be nearly zero. 
The low $T$ behavior is enlarged in Fig. 1(b). In this figure, $T_{SP}$ is observed to decrease slightly with increasing $H$ as indicated by solid arrows. Note that a hysteresis between the $T$ up and $T$ down processes of $\chi$($T$) is detected in the field range for 0.5 T $\le$ $H$ $\le$ 1 T as is shown by open arrows in Fig.~\ref{f1}(b). Here, $\chi$($T$) is recorded in the zero-field cooling process where $H$ is applied at the lowest $T$ and the field cooling where $H$ is applied for $T$ $>$ $T_{SP}$.

\begin{figure}
\includegraphics[width=60mm]{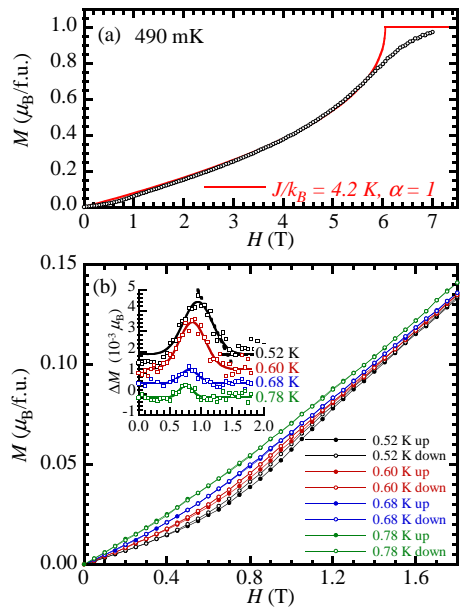}
\caption{\label{f2} (a) $M$($H$) curve at 0.49 K. The solid curve is the theoretical calculation for $\alpha$=1 with $J$/$k_{B}$=4.2 K at $T$=0 K.
(b) Hysteresis in the $M$($H$) curve between $H$ increase and decrease processes. The difference $\Delta$$M$, which is fitted by a Gaussian, is shown in the inset.
}
\end{figure}

\begin{figure}
\includegraphics[width=60mm]{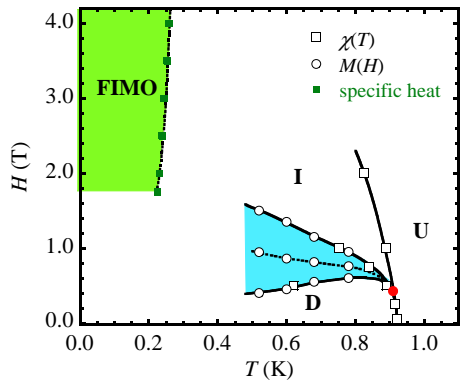}
\caption{\label{f3} $H$-$T$ phase diagram of D-F$_{5}$PNN. The center and width of the hysteresis for the D-I boundary region are assigned as the peak position and full width of the Gaussian curve fitted to $\Delta$$M$ indicated in the inset of Fig.~\ref{f2}. The solid (red) circle represents a Lifshitz point. The FIMO phase determined by previous specific heat measurements, which were the subject of some of our unpublished work, is also shown here, for clarity.
}
\end{figure}

The $M$($H$) curve recorded during a field-up ramp at 0.49 K is shown in Fig.~\ref{f2}(a).
In the field range of 2$\sim$6 T, the $M$($H$) curve can be reproduced by the theoretical calculation for the $S$=1/2 Heisenberg linear chain at $T$=0 K using the same parameters of $J$/k$_{B}$=4.2 K and $g$ = 2.006 as is represented by the solid line\cite{griffiths}. $M$($H$) finally reaches the saturation value of 1 $\mu_{B}$ at around 7 T. 
The hysteresis is also observed in the $M$($H$) curves below $T_{SP}$ as shown in Fig~\ref{f2}(b). The inset of this figure represents the difference of magnetization $\Delta$$M$ between $H$ up and down traces, showing the $T$ dependence. 
To estimate the field range of $\Delta$$M$, we have fitted $\Delta$$M$ with a Gaussian curve, plotted by the solid line in the inset of Fig~\ref{f2}(b). 
The observed magnetization results are summarized in the $H$-$T$ phase diagram shown in Fig.~\ref{f3}, in which the FIMO phase determined by previous specific heat measurements is combined.
The entire profile of the $H$-$T$ phase diagram in Fig.~\ref{f3} is consistent with that proposed by the theoretical study, demonstrating that D-F$_{5}$PNN is an ideal compound for studying the SP transition\cite{h-t}. It is noted here that no thermal hysteresis is seen at the U-D boundary, while first-order character is reported by neutron scattering studies. Such a discrepancy is simply attributed to the sample size. Including F$_{5}$PNN, the large crystal seems to show a first order character in the transition\cite{d-f5pnn1, *d-f5pnn2}. Therefore, in this work, we have used a small and thin crystal. 

The second order U-D line intercepts the horizontal axis at $T_{SP}$(0)=0.92 K. Along the D-I line, the onset of hysteresis is observed at the Lifshitz point ($H^{*}$, $T^{*}$)=(0.43 T, 0.91 K) as represented by the solid circle in this figure. These values give $\mu_{B}$$H^{*}$/$k_{B}$$T_{SP}$(0)=0.31, which is close to the predicted value of 0.28 by the NF theory\cite{nf}. By extrapolating the D-I phase boundary to the low $T$ side, the upper field critical line seems to terminate the FIMO phase. The shape of the present FIMO phase is shown to be markedly different from the semicircular shape of conventional dimer systems. Actually, in the FIMO phase for the present compound, no Bragg peak due to commensurate AF ordering has been detected in neutron scattering measurements. 
From the observations detailed above, we can conclude that the incommensurate structure is realized in I-phase for low-$T$ and high-$H$ region.

\begin{figure}
\includegraphics[width=60mm]{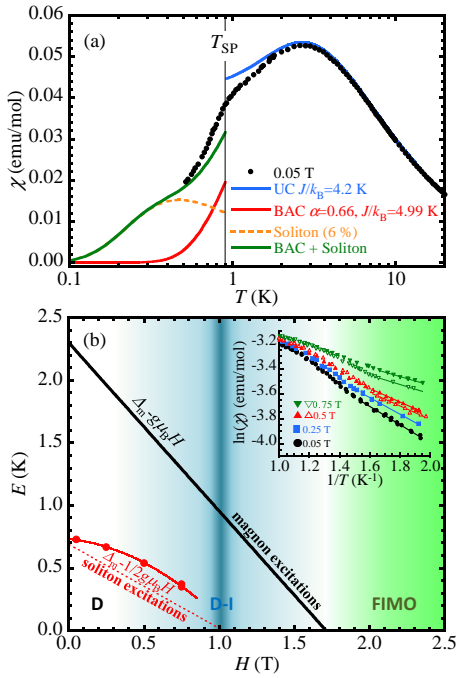}
\caption{\label{f4} (a) Magnon and soliton contributions to $\chi$($T$) below $T_{SP}$. The UC and BAC correspond to the theoretical calculations for an $S$ = 1/2 AF Heisenberg linear chain model and an $S$ = 1/2 AF Heisenberg bond alternation chain model, respectively. 
(b) Spin excitations under an external magnetic field. Only the lowest branch is shown for both the soliton (doublet) and the magnon (triplet) excitations. The colored areas stand schematically for the D-I boundary and the FIMO phase. Inset shows the gap estimation procedure from the ln($\chi$) vs. 1/$T$ plot.
}
\end{figure}

Once again, we focus on the low-$T$ behavior of $\chi$($T$). According to the inelastic NS measurements, the magnon gap $\Delta_{m}$ is estimated to be 0.2 meV (2.3 K or 1.7 T) from the dispersion relation of the triplet magnon branch with $J$/$k_{B}$ = 4.99 K and bond alternation parameter $\alpha$ = 0.66 at $H$ = 0 T\cite{d-f5pnn-nd}. It is evident that these parameters do not reproduce $\chi$($T$) below $T_{SP}$ as is shown in Fig.~\ref{f4}(a). 
Hence, we must take account of the soliton excitations.
In order to understand this situation, we estimate an effective spin gap $\tilde{\Delta}$ from the low-$T$ part of the observed $\chi$($T$). Plotting ln($\chi$($T$)) vs. 1/$T$ at each $H$ yields an $H$-dependence of the effective gap $\tilde{\Delta}$ ($H$) as shown in Fig.~\ref{f4}(b). $\tilde{\Delta}$($H$) has a finite curvature and does not follow a straight line. This is the case due to a combination of two things. The first is that the $T$ range is not sufficiently low for the gap estimation in high-$H$, where the gap becomes small. The second is that $\chi$($T$) contains both contributions from magnon and soliton excitations below $T_{SP}$. The extrapolated zero field-gap $\tilde{\Delta}$(0) = 0.73 K is in good agreement with the soliton gap $\Delta_{s}$ deduced from the relation $\Delta_{s}$ $\sim$ 0.3$\Delta_{m}$, where the magnon gap $\Delta_{m}$=2.3 K\cite{nf, h-t}. The agreement indicates that $\chi$($T$) is dominated by soliton excitations in the low-$T$ and the low-$H$ region. Accordingly, we can evaluate the contribution of soliton excitations to $\chi$($T$) separately using the value $\tilde{\Delta}$(0) = 0.73 K. For simplicity, we consider the soliton excitations as an ensemble of isolated doublet excitation. Thus $\chi$($T$) for soliton excitations is calculated as the expectation value of magnetization for free $S$=1/2 spins with a finite excitation gap $\Delta_{s}$. The calculated curve is then compared with the observed $\chi$($T$) at $H$=0.05 T as is shown in Fig.~\ref{f4}(a). Good reproducibility is obtained by the combination of magnon and soliton susceptibilities below $T_{SP}$, if we assume the 6 $\%$ of soliton density, namely $S$=1/2 spin per 17 lattice sites, as shown in Fig.~\ref{f4}(a). The value of $\Delta_{m}$ is nearly identical with the critical field value where the FIMO phase starts to appear, while $\Delta_{s}$ coincides with the center field of the D-I phase as represented in Fig.~\ref{f4}(b). The coincidence of these values is not accidental, but reflects the essential feature of an SP transition. From the observed experimental facts, we conclude that magnon and soliton excitations are responsible for the FIMO phase and the D-I transition, respectively. 

\begin{figure}
\includegraphics[width=60mm]{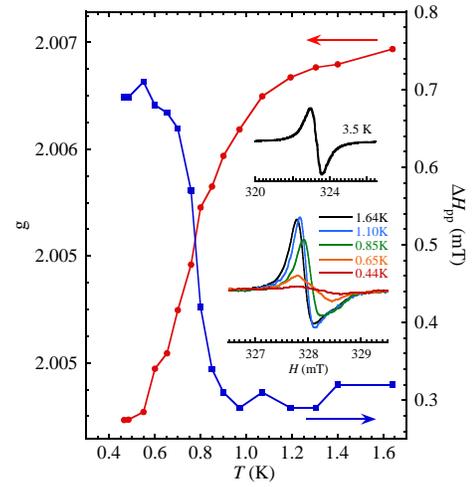}
\caption{\label{f5} $T$-dependence of $g$ and $\Delta$$H_{pp}$ at around $T_{SP}$. The inset shows typical ESR spectra.
}
\end{figure}

In the final part of this paper, we check the validity of our interpretation based on the soliton excitations in comparison with the result of ESR spectroscopy. The observed $T$ dependence of the $g$-value and the linewidth $\Delta$$H_{pp}$ is shown in Fig.~\ref{f5}.
Both quantities show a clear change in behavior at around $T_{SP}$. When the temperature is decreased, $\Delta$$H_{pp}$ increases and the $g$-value approaches an isotropic value of 2.0. These aspects can be understood qualitatively by considering the dilution of an exchange narrowing effect and suppression of the magnetic contributions that give rise to a $g$-shift, respectively. In addition, the change in the line profile is also informative. As shown in the inset of Fig.~\ref{f5}, the line profile is slightly asymmetric above 0.7 K. In contrast, a symmetric derivative curve is obtained at the low-$T$ range below 0.7 K in addition to at the high-$T$ range above 3.5 K. Unfortunately, we have no ESR data between 1.64 K and 3.5 K, due to bubbling of the liquid $^{4}$He.

To explain the observed results, we consider the following scenario. At low $T$, the symmetric ESR mainly comes from soliton excitations, as the spin susceptibility of excited magnons is nearly zero as shown Fig.~\ref{f4}(b). In the middle-$T$ range, the magnon contribution is developed, resulting in the asymmetric line shape to the ESR absorption curve. Because slightly different $g$-values are expected between dispersive magnon and localized soliton, their superposition can give rise to an asymmetric line shape. Such a picture with elementary excitations is no longer relevant in the U-phase, so that the ESR spectrum becomes symmetric as we have observed above 3.5 K. Even above $T_{SP}$, the asymmetric ESR line shape is held at least up to 1.64 K, but below 3.5 K. This means that a pseudo-gap opens between 1.64 K and 3.5 K and is consistent with the $\chi$($T$) that starts to deviate from the UC line below this $T$ range. 

In summary, we have performed magnetization and ESR measurements on D-F$_{5}$PNN. The resultant $H$-$T$ phase diagram consists of U, D, and I phases with a hysteresis region that starts from the Lifshitz point into the D-I phase boundary. 
The susceptibility in the D-phase is well reproduced by taking contributions from both magnon and soliton excitations into account. This interpretation is consistent with the temperature-dependent behaviors of ESR. Based on these experimental facts, we conclude that D-F$_{5}$PNN is the most ideal compound for studying SP transition found so far.

\begin{acknowledgments}
The authors thank M. Matsumoto and T. Matsushita for stimulating discussions. This work was partially supported by a Grant-in-Aid for Scientific Research, No.15K06424, No.25220605 and No.25287076.

\end{acknowledgments}
\bibliography{D-F5PNN-Inagaki}
\end{document}